\documentclass[aps,preprint,superscriptaddress,longbibliography]{revtex4-1}

\usepackage{graphicx}
\usepackage{dcolumn}
\usepackage{bm}
\usepackage{amssymb}
\usepackage{microtype}
\usepackage{xfrac}
\usepackage{array}
\usepackage{gensymb}
\usepackage{xcolor}
\usepackage[normalem]{ulem}

\newcommand{\pp}{PtPb$_4$}
\newcommand{\g}{$\Gamma$}
\newcommand{\gx}{$\Gamma$-X}
\newcommand{\ef}{\textrm{$E_F$}}

\newcommand{\kz}{$k_z$}
\newcommand{\kx}{$k_x$}
\newcommand{\ky}{$k_y$}

\newcommand{\hl}[1]{\textcolor{black}{#1}}

\begin{document}

\title{Nonsymmorphic Symmetry-Protected Band Crossings in a \hl{Square-Net} Metal PtPb$_4$}


\author{Han Wu}
\thanks{These authors contributed equally.}
\affiliation{Department of Physics and Astronomy and Rice Center for Quantum Materials, Rice University, Houston, TX, 77005 USA}

\author{Alannah~M.~Hallas}
\thanks{These authors contributed equally.}
\affiliation{Department of Physics and Astronomy and Rice Center for Quantum Materials, Rice University, Houston, TX, 77005 USA}
\affiliation{Department of Physics and Astronomy and Quantum Matter Institute, University of British Columbia, Vancouver, British Columbia V6T 1Z1, Canada}

\author{Xiaochan Cai}
\thanks{These authors contributed equally.}
\affiliation{School of Physical Science and Technology, ShanghaiTech University, Shanghai 201210, China}

\author{Jianwei Huang}
\affiliation{Department of Physics and Astronomy and Rice Center for Quantum Materials, Rice University, Houston, TX, 77005 USA}

\author{Ji Seop Oh}
\affiliation{Department of Physics, University of California, Berkeley, Berkeley, California 94720, USA}
\affiliation{Department of Physics and Astronomy and Rice Center for Quantum Materials, Rice University, Houston, TX, 77005 USA}

\author{Vaideesh Loganathan}
\affiliation{Department of Physics and Astronomy and Rice Center for Quantum Materials, Rice University, Houston, TX, 77005 USA}

\author{Ashley~Weiland}
\affiliation{Department of Chemistry \& Biochemistry, University of Texas at Dallas,
Richardson, Texas 75080, United States}

\author{Gregory~T.~McCandless}
\affiliation{Department of Chemistry \& Biochemistry, University of Texas at Dallas,
Richardson, Texas 75080, United States}
\affiliation{Department of Chemistry and Biochemistry, Baylor University, Waco, Texas 76798, United States}

\author{Julia~Y.~Chan}
\affiliation{Department of Chemistry \& Biochemistry, University of Texas at Dallas,
Richardson, Texas 75080, United States}
\affiliation{Department of Chemistry and Biochemistry, Baylor University, Waco, Texas 76798, United States}

\author{Sung-Kwan Mo}
\affiliation{Advanced Light Source, Lawrence Berkeley National Laboratory, Berkeley, CA 94720, USA}

\author{Donghui Lu}
\affiliation{Stanford Synchrotron Radiation Lightsource, SLAC National Accelerator Laboratory, Menlo Park, California 94025, USA}

\author{Makoto Hashimoto}
\affiliation{Stanford Synchrotron Radiation Lightsource, SLAC National Accelerator Laboratory, Menlo Park, California 94025, USA}

\author{Jonathan Denlinger}
\affiliation{Advanced Light Source, Lawrence Berkeley National Laboratory, Berkeley, CA 94720, USA}

\author{Robert J. Birgeneau}
\affiliation{Department of Physics, University of California, Berkeley, Berkeley, California 94720, USA}
\affiliation{Materials Sciences Division, Lawrence Berkeley National Laboratory, Berkeley, California 94720, USA}
\affiliation{Department of Materials Science and Engineering, University of California, Berkeley, USA}

\author{Andriy H. Nevidomskyy}
\affiliation{Department of Physics and Astronomy and Rice Center for Quantum Materials, Rice University, Houston, TX, 77005 USA}

\author{Gang Li}
\email{ligang@shanghaitech.edu.cn}
\affiliation{School of Physical Science and Technology, ShanghaiTech University, Shanghai 201210, China}
\affiliation{ShanghaiTech Laboratory for Topological Physics, ShanghaiTech University, Shanghai 201210, China}

\author{Emilia Morosan}
\email{emorosan@rice.edu}
\affiliation{Department of Physics and Astronomy and Rice Center for Quantum Materials, Rice University, Houston, TX, 77005 USA}

\author{Ming Yi}
\email{mingyi@rice.edu}
\affiliation{Department of Physics and Astronomy and Rice Center for Quantum Materials, Rice University, Houston, TX, 77005 USA}

\date{\today}

\maketitle

\section{Abstract}

Topological semimetals with symmetry-protected band crossings have emerged as a rich landscape to explore intriguing electronic phenomena. Nonsymmorphic symmetries in particular have been shown to play an important role in protecting the crossings along a line (rather than a point) in momentum space. Here we report experimental and theoretical evidence for Dirac nodal line crossings along the Brillouin zone boundaries in PtPb$_4$, arising from the nonsymmorphic symmetry of its crystal structure. Interestingly, while the nodal lines would remain gapless in the absence of spin-orbit coupling (SOC), the SOC in this case plays a detrimental role to topology by lifting the band degeneracy everywhere except at a set of isolated points. Nevertheless, \hl{
the nodal line is observed to have a bandwidth much smaller than that found in density functional theory (DFT). Our findings reveal PtPb$_4$ to be a  material system with narrow crossings approximately protected by non-symmorhpic crystalline symmetries.}


{\bf{Keywords:}} Topology, nonsymmorphic symmetry, electron correlations, angle-resolved photoemission spectroscopy, dynamical mean field theory
\clearpage

\section{Introduction}

Since the discovery of topological insulators more than a decade ago, the classification of quantum materials has undergone a revolution, where quantum materials are now categorized by their topological properties and associated symmetries~\cite{Hasan2010TI,Qi2011,Zhang2009,Xia2009,Chen178,Tang2019,Zhang2019,Vergniory2019,Kruthoff2017,Slager2013}. This is epitomized by the study of topological semimetals (TSMs), which are 3-dimensional analogs to graphene where the bulk bands cross without opening an energy gap~\cite{Armitage2018}. 
Crystalline symmetries, such as nonsymmorphic symmetry, can play a crucial role in protecting the band crossings along a continuous line or loop in momentum space in TSMs, yielding what is known as a nodal line semimetal (NLS). Nonsymmorphic symmetries combine a fractional lattice translation with either a mirror reflection (glide plane) or a rotation (screw axis), resulting in a band-folding with crossings at the Brillouin zone (BZ) boundaries that are protected against hybridization~\cite{wang2016hourglass,yang2017symmetry}. As long as the nonsymmorphic symmetry remains intact, these so-called essential band crossings are impervious to the presence of SOC~\cite{wang2016hourglass,yang2017symmetry, bzduvsek2016nodal}.
\hl{If furthermore the degenerate bands only slightly disperse, one observes the symmetry-protected crossings of narrow bands, which when doped to the chemical potential may host correlated topological phases~\cite{Bistritzer12233, TBG-Nature1, TBG-Nature2}. }

In contrast to the vast repository of materials with band crossings that are unprotected against SOC, there are only a handful of materials that realize nonsymmorphic symmetry-protected Dirac crossings. The search for nonsymmorphic topological materials has largely been guided by a work from Young and Kane~\cite{Young2015}, which demonstrated that two-dimensional square net motifs can generate Dirac nodes, when the square net is itself embedded in a unit cell that is twice as large and hence the two atoms in the unit cell are related by a glide plane. This blueprint has been followed in the case of ZrSiS, which has the nonsymmorphic space group $P4/nmm$ with Si occupying a square net. ARPES measurements on ZrSiS~\cite{Schoop2016,neupane2016observation} and several isosructural compounds~\cite{takane2016dirac,schoop2018tunable} have revealed nonsymmorphic symmetry-protected Dirac nodal lines with linear dispersions over more than 2~eV. Several years onward, ZrSiS and its structural analogs, whose electronic structure can be well captured by DFT, remain one of the few experimental manifestations of this class of topological materials~\cite{klemenz2019topological,klemenz2020systematic}.

Here we report the discovery of a nonsymmorphic symmetry protected topological semimetal \hl{displaying narrow bands along the BZ boundary, \pp.} 
Distinct from a previous report on PtPb$_4$ crystals that exhibit a crystal structure consistent with space group $Ccce$~\cite{lee2020evidence}, our post-annealed crystals exhibit tetragonal symmetry of the space group $P4/nbm$. Our ARPES measurements reveal a set of nearly flat degenerate bands along the BZ boundary that appear to originate from the linear crossing of two bulk bands along the orthogonal momentum direction. Careful analysis of space group symmetries and their representations in the BZ shows that the band degeneracies along the BZ boundary originate from nonsymmorphic symmetry elements. 
Surprisingly, despite the large SOC expected from the heavy elements constituting \pp, 
the observed splitting of the bands is much smaller than predicted by \textit{ab initio} calculations based on DFT. 
%
%
By comparing our ARPES measured bands with those from both DFT and dynamic mean field theory (DMFT) calculations~\cite{PhysRevLett.62.324, Hartmann-1989, Metzner-1989, RevModPhys.68.13}, we further explore the role of  electron correlation effects on the band details. \pp~therefore \hl{is one of the few reported nonsymmorphic symmetry-protected TSMs outside the well-known ZrSiS family that exhibits narrow bands  in the presence of non-trivial topology.}

\section{Results and Discussion}

\subsection{Crystal structure and nonsymmorphic symmetry}
PtPb$_4$, grown by the metallic flux method, forms in the tetragonal space group 125 ($P4/nbm$) with lattice parameters a = 6.66~\AA~and c = 5.978~\AA~\cite{Rosler_1951}. No structural phase transitions are found down to 2 K. A recent study on PtPb$_4$ suggested that this material is polymorphic, with similar formation energies for tetragonal and orthorhombic structures~\cite{lee2020evidence}. In our study, we determined that extensive post-growth annealing (described in Supplementary Note 1 and Note 2) was critical to obtaining a single phase tetragonal material. The tetragonal crystal structure of \pp~consists of staggered layers of Pt and Pb, as seen in Fig. \ref{fig:fig1}a. Two layers of Pb (gold) are sandwiched between consecutive layers of Pt (blue). Each Pb layer forms a Shastry--Sutherland lattice while a square net is formed by each Pt layer. This crystal structure exhibits nonsymmorphic symmetry through a glide-mirror operation, as illustrated in Fig.~\ref{fig:fig1}a. For the Pb in site A, a mirror reflection operation $m_z$ brings it to site B, which is not an allowed position in the structure. An additional fractional translation is needed ($t: \frac{1}{2}, \frac{1}{2}, 0$) to bring it to the allowed atomic site C. We note that \pp~is structurally similar but not isostructural with PtSn$_4$, which has been reported to exhibit Dirac nodal arc surface states~\cite{wu2016dirac}, features that are apparently unrelated to the nonsymmorphic symmetry. 

\subsection{Electronic structure and evidence for nodal line} 
\begin{figure}[]
\includegraphics[width=\textwidth]{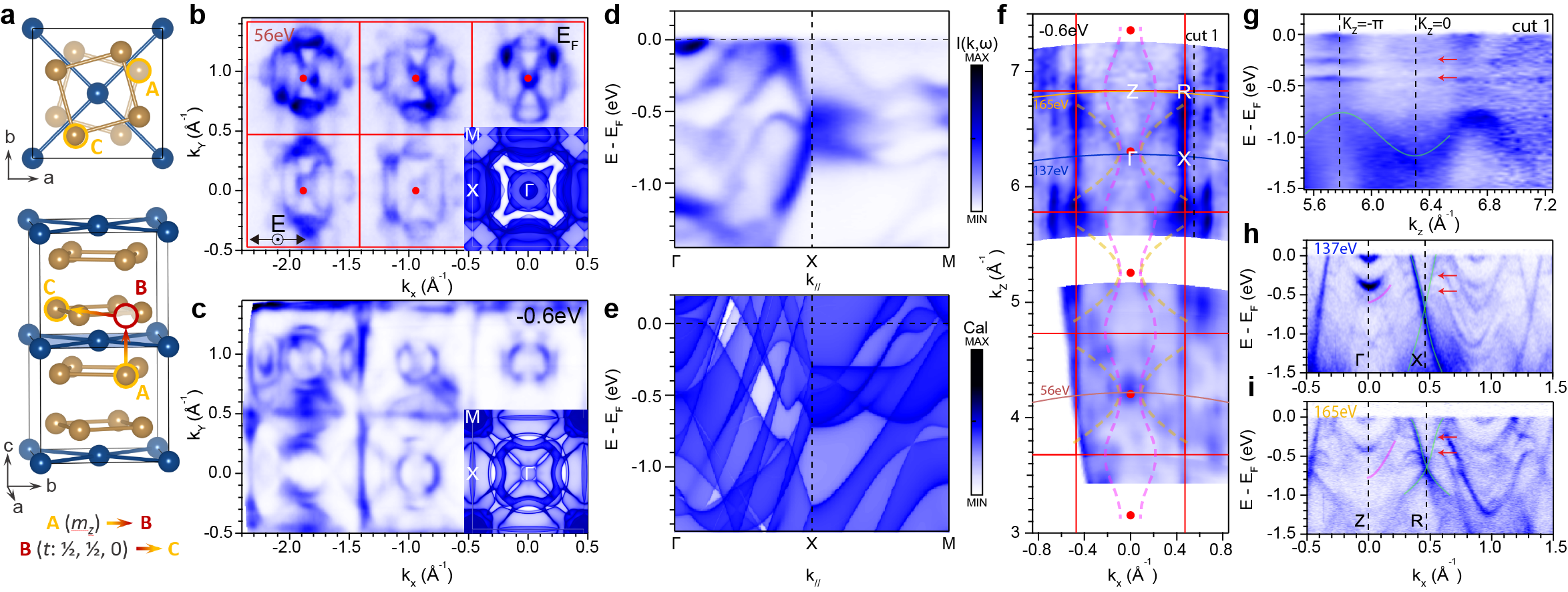} 
\caption{\textbf{Crystal structure and electronic structure of \pp.}
(a) Top view and side view of the crystal structure of \pp~(Pt: blue; Pb: gold). Illustration of a non-symmorphic group operation is indicated at the bottom that includes a mirror reflection and a translation. (b)	Measured Fermi surface in the \kx-\ky~plane integrated within 10 meV of the Fermi level. Polarization vector is as shown. DFT slab-calculation is shown for comparison. (c) ARPES constant energy contour at -0.6 eV below the Fermi level. Corresponding slab calculation is shown. (d) Band dispersions measured along the $\Gamma$-X-M direction. (e) Calculated \kz-integrated dispersions along $\Gamma$-X-M from slab calculation. Only bulk states are shown. Spin-orbit coupling (SOC) is not included in this set of calculations. (f) Constant energy contour taken at -0.6eV from a photon energy dependence study. An inner potential of 17eV was used. Selected bulk bands that are dispersive along k$_z$ are marked. (g) A dispersion cut along cut 1 shown in (f). (h)-(i) Measured dispersions taken at 137eV and 165eV, respectively. Data in (b),(c) are taken at 25 K. Data in (d) are taken at 100 K.}
\label{fig:fig1}
\end{figure}

We investigated the electronic structure of \pp~via ARPES measurements. The Fermi surfaces (FS) measured under an in-plane polarization along the horizontal direction are shown in  Fig.~\ref{fig:fig1}b, which consists of flower-like Fermi pockets centered at the BZ center. We note that the intensity of the FS appears to indicate C$_4$ symmetry-breaking, which has also been reported by a laser-ARPES work~\cite{lee2020evidence}. However, we caution that since our polarization used is along the k$_x$ direction, the photoemission matrix elements also break C$_4$ symmetry, which prevent us from concluding whether the intrinsic electronic structure breaks C$_4$ symmetry. However, our powder x-ray diffraction measurements show no evidence of C$_2$ symmetry in our annealed crystals, suggesting that our bulk bands should exhibit C$_4$ symmetry. Notably, at a binding energy of 0.6 eV below the Fermi level (\ef), an intense grid-like feature appears along the lines that coincide with the boundaries of the BZ, as evident in the measured constant energy contour (Fig.~\ref{fig:fig1}c), reminiscent of the nonsymmorphic symmetry-protected features originally predicted by Young and Kane~\cite{Young2015}. To see the band dispersions that give rise to this feature, we show the spectral images measured along the high symmetry directions of the BZ (Fig.~\ref{fig:fig1}d). First, along the \gx~direction, a series of electron bands appear. Closer to the X point, we also observe two highly dispersive bands that meet at the X point. In contrast to the \g-X direction, intensity from dispersions along the X-M direction are mostly confined within the energy window of $-0.9$~eV to $-0.3$~eV, where a set of largely flat bands appear, in particular, near $-0.6$~eV. These are the bands that give rise to the grid-like features outlining the BZ boundaries in the constant energy contour in Fig.~\ref{fig:fig1}c. To identify whether the crossing is of bulk or surface nature, we carried out a photon energy dependence study for probing along the k$_z$ direction. A number of features in the constant energy contour taken at -0.6eV are shown to be periodic, and therefore identified as bulk bands  (Fig.~\ref{fig:fig1}f). In particular, we find that the bands giving rise to the Dirac crossing at the BZ boundary exhibit different band velocities at k$_z$=0 and $\pi$. This is illustrated in a cut near the X-R direction (Fig.~\ref{fig:fig1}g), where the Dirac dispersion is observed to be periodic along k$_z$, indicating that it is bulk in nature. The grid-like feature this band forms in the constant energy contour indeed appears at all BZ boundaries, respecting the bulk C$_4$ symmetry. Moreover, we also note a clear k$_z$ broadening effect for this band, seen in the broadened intensity where this band disperses across k$_z$. This is an effective integration along k$_z$ due to the low resolution of the photoemission process in the out-of-plane direction~\cite{Strocov2003}. We carried out semi-infinite slab calculations using DFT based on the Green's function method~\cite{Sancho1985}. The resulting calculated FS, shown in the bottom right panel of Fig.~\ref{fig:fig1}b, reproduces the series of pockets centered at \g~seen in ARPES. At $-0.6$~eV, large pockets centered at \g~are also reproduced. Importantly, this calculation also reproduces the grid like outline of the BZ boundaries (bottom right panel in Fig.~\ref{fig:fig1}c). Due to the strong k$_z$ broadening observed, we compared the measured dispersions with the calculated \kz~integrated band dispersions along \gx. Largely dispersive features appear to qualitatively match those seen in the ARPES data. Along the high-symmetry X-M line, a number of bands appear in the energy range centered at $-0.6$~eV, which likely correspond to the bands giving rise to the grid-like feature in the constant energy contour. However, the calculated dispersions along the X-M line span a larger energy window than what is observed experimentally, \hl{indicating a subtle mechanism that is not captured by the DFT calculations.}

\begin{figure}[]
\includegraphics[width=\textwidth]{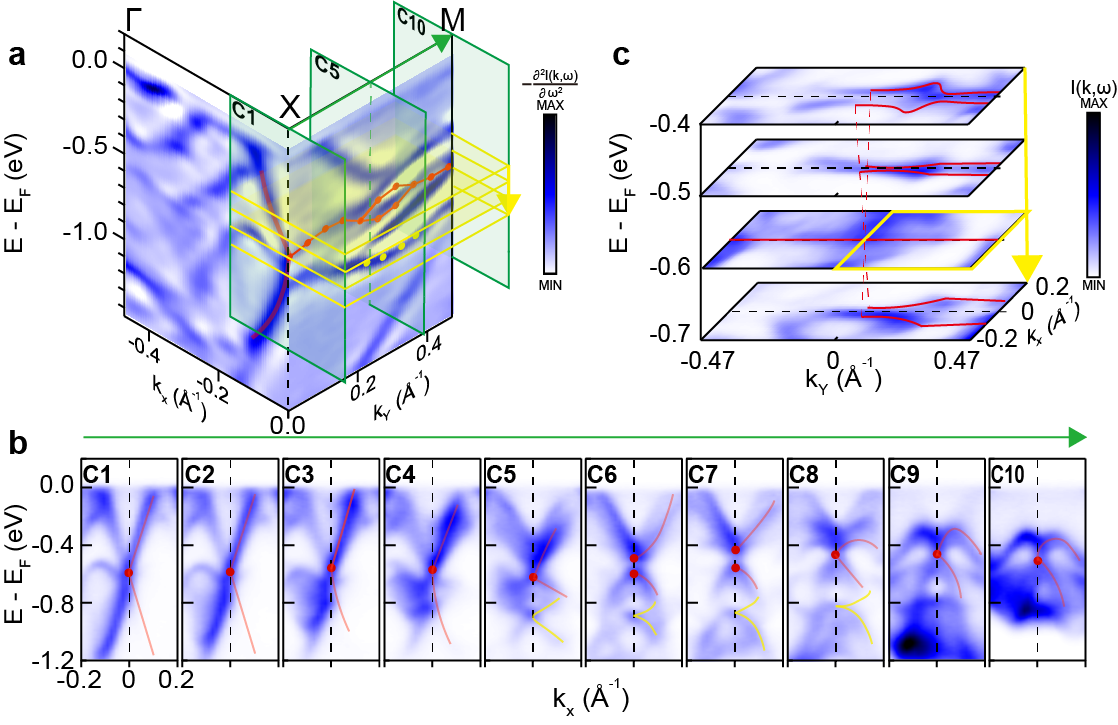} 
\caption{\textbf{Dirac crossings along the nearly flat nodal line.}
(a) 2D curvature of measured spectral image along the $\Gamma$-X-M direction. Red lines along \gx~are guides to the eye of the dispersive band for the Dirac crossing at X. Markers along X-M indicate the fitted positions of the band crossings from energy distribution curves. (b) Dispersions measured along the direction orthogonal to X-M as shown by green slices in (a). Band crossings with unresolvable gaps are indicated by red arrows while gapped crossings are indicated by green and blue arrows. A second set of Dirac crossings are marked by yellow lines, where the crossing energies are marked as yellow markers in panel (a). (c) Constant energy contours as indicated in panel (a) along X-M to show the evolution of the crossings leading to the nodal line around -0.6 eV. Data is taken at 100 K. The polarization used is the same as in Fig.~\ref{fig:fig1}(d).
}
\label{fig:fig2}
\end{figure}

In order to shed more light on the nature of the band crossings along the edges of the BZ, we analyze in detail the measured band dispersions along and perpendicular to the X-M BZ boundary (Fig.~\ref{fig:fig2}). To better visualize the band dispersions, we plot the 2D curvature of the spectral image along the \g-X-M path in Fig.~\ref{fig:fig2}a. The pair of bands highlighted in red in Fig.~\ref{fig:fig2}a disperse along \g-X to meet at a degenerate point at X, then remain nearly degenerate across the X-M edge of the zone. 
To see this, we examine a series of ten cuts (C1 to C10) perpendicular to the X-M direction (Fig.~\ref{fig:fig2}b). From C1, the linear Dirac crossing can be clearly observed near $-0.6$~eV. As we move away from the X point, this Dirac crossing remains centered at nearly the same energy. 
In C6 and C7, a gap is resolved at the Dirac point, and subsequently closes approaching the M point (C8 to C10), although a small gap may persist along the cut beyond our experimental resolution. \hl{We note that the bandwidth of this nodal line is less than 0.2~eV along the BZ boundary direction.}
In addition, another set of nearly degenerate bands can be observed near $-0.9$~eV, as marked in yellow on cuts C5 to C8. The crossing points of these bands are also indicated for clarity by yellow dots along the X-M line in Fig.~\ref{fig:fig2}a. 
We therefore conclude that what appears to be a nearly flat grid-like feature at $-0.6$~eV in the ARPES data is actually a set of two bands that cross or nearly cross along the X-M line at the BZ edge. This can also be seen in constant energy contours. As shown in Fig.~\ref{fig:fig2}c, a line segment lining the BZ boundary at $-0.6$~eV evolves into two separate features both above and below in energy due to the Dirac bands dispersing away from the nodal crossing. 

\subsection{Symmetry analysis and first-principle calculations}
\begin{figure}[]
\includegraphics[width=\textwidth]{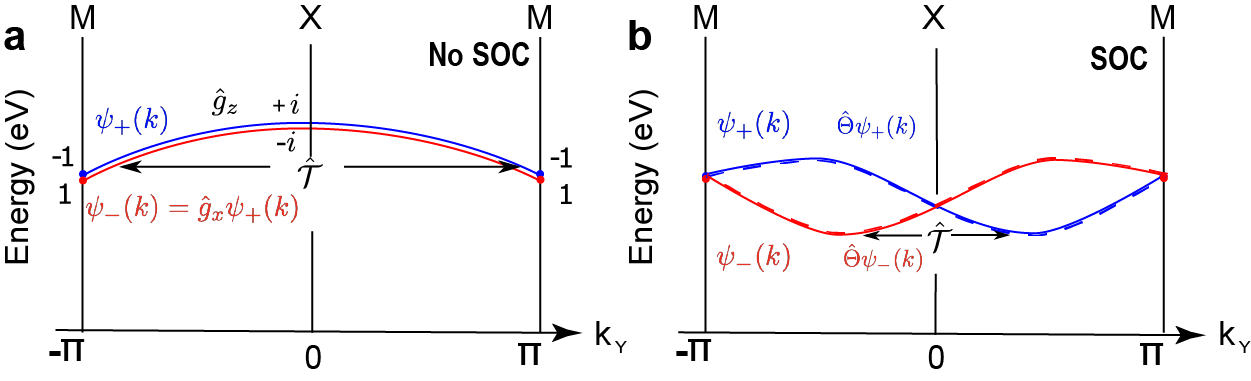} 
\caption{\textbf{Nonsymmorphic symmetry-protected band degeneracy.} (a) Without SOC, glide symmetries $\hat{g}_{z}$ and $\hat{g}_{x}$ guarantee the degeneracy of the two bands $\psi_{+}$ and $\psi_{-}=\hat{g}_{x}\psi_{+}$. Here the blue and red solid lines denote the two bands with $\pm$ $\hat{g}_{z}$ eigenvalues $\pm ie^{-\mathbf{k}_{y}/2}$. Furthermore, These two bands  become symmetric with respect to $\mathbf{k_{y}}=0$ under time reversal symmetry ($\hat{\cal T}$). (b) With SOC, the blue and red bands have to separate and each of them becomes doubly degenerate due to the Kramer's pairing. The presence of both time reversal $\hat{\cal T}$ and parity $\hat{\cal P}$ symmetries enforces the four-fold band crossings at $\mathbf{k_{y}}=0, \pm\pi$ (X and M). 
}
\label{fig:fig3}
\end{figure}

The intriguing features in both the measured and calculated band dispersions prompted us to examine these findings in the context of the nonsymmorphic symmetry of \pp. The glide mirror symmetries of the space group can be represented by $\hat{g}_{x}=\{m_{010}|\frac{1}{2},0,0\}$ and $\hat{g}_{z}=\{m_{001}|\frac{1}{2},\frac{1}{2},0\}$ (Note that these operations are defined with respect to the standard primitive cell for space group 125, which has its origin shifted by (0, $\sfrac{1}{4}$, 0) with respect to Fig.~\ref{fig:fig1}(a)). As these operations are orthogonal to each other (anticommuting), in the absence of SOC, the bands $\psi_{+}$ and $\psi_{-}=\hat{g}_{x}\psi_{+}$ carry opposite eigenvalues of $\hat{g}_{z}$ and are hence degenerate along the BZ boundary (Fig.~\ref{fig:fig3}a). With the inclusion of SOC, a gap opens along the BZ boundary except at the highest symmetry points, owing to the breaking of the anti-commutation relation between $\hat{g}_{x}$ and $\hat{g}_{z}$ (Fig.~\ref{fig:fig3}b). With SOC present, the double space-group operation requires $\hat{g}_{x}$ and $\hat{g}_{z}$ to \textit{commute} in spin space as well. Thus, instead of the anti-commutator $\{\hat{g}_{x}, \hat{g}_{z}\}=0$, one has the commuting relation $[\hat{g}_{x}, \hat{g}_{z}]=0$ under SOC.
As a result, the action of the glide plane on a band $\psi_{-}=\hat{g}_{x}\psi_{+}$ is no longer degenerate with $\psi_{+}$ as they now carry the same $\hat{g}_{z}$ eigenvalue. 
Consequently, every band along X-M (and the equivalent direction R-A) is only degenerate with its Kramers' pair, except at the high symmetry points at $k_y = 0$ (X) or $\pi$ (M-point), where the four-fold degeneracy remains due to the presence of both time reversal $\hat{\cal T}$ and parity $\hat{\cal P}$ symmetries. We therefore find that, in the absence of SOC, the nonsymmorphic symmetry in \pp~must protect the band degeneracy along the BZ boundary, similar to earlier theoretical reports on nonsymmorphic structures~\cite{Yang2019,yang2017symmetry}. However, the presence of SOC lifts the degeneracy except at a set of isolated points. The nonsymmorhpic symmetry of the crystal could also explain the strong intensity asymmetry of the Dirac dispersions about the BZ boundary (Fig.~\ref{fig:fig1}h-i), where the glide mirror symmetry switches the parity of the orbital symmetries, as has been observed in the iron pnictides~\cite{Brouet2012}.

\begin{figure}[]
\includegraphics[width=\textwidth]{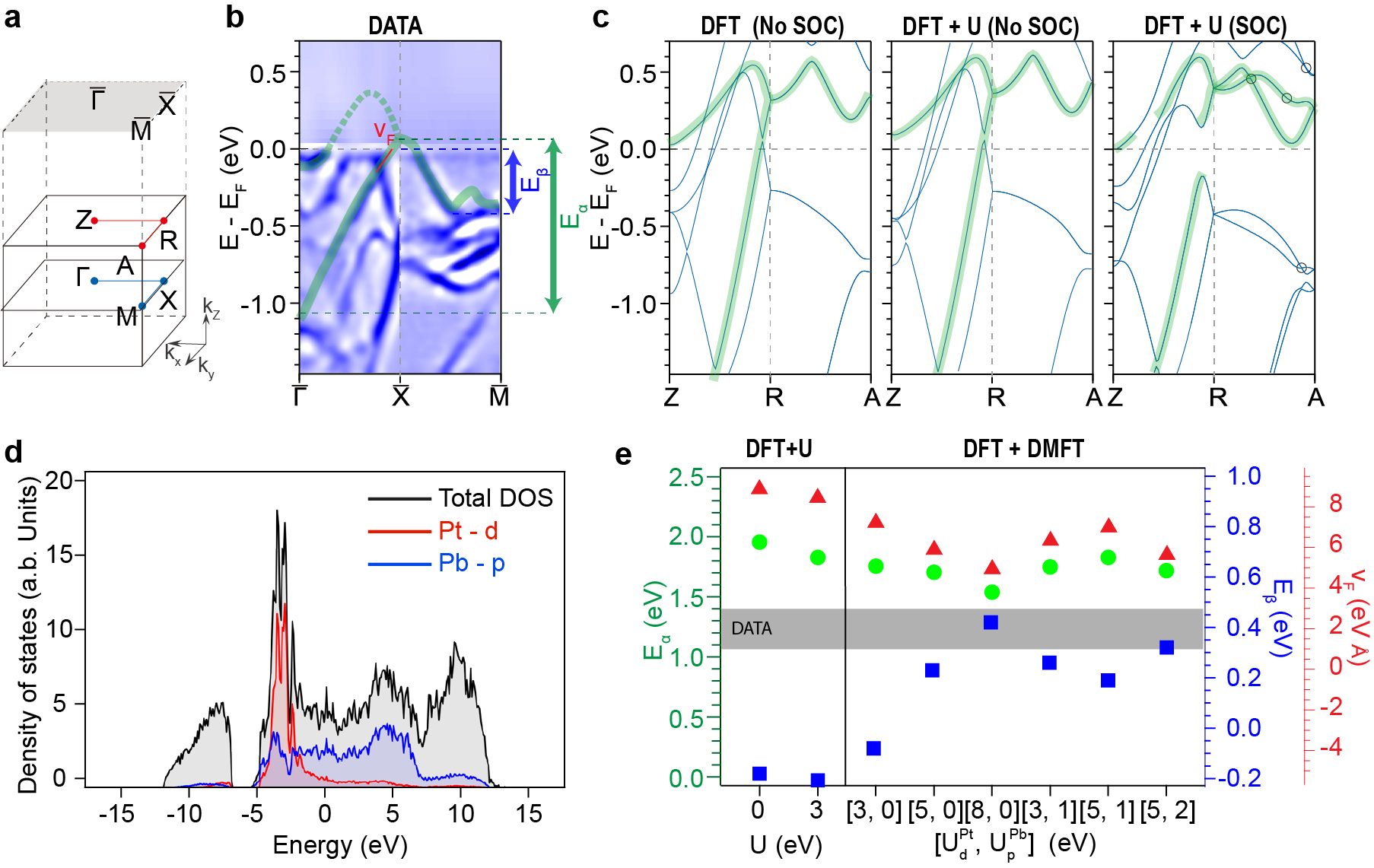} 
\caption{\textbf{First-principle calculations of the electronic structure.} (a) BZ notations for the tetragonal unit cell and the projected surface BZ. (b) 2D curvature of measured spectral image along $\overline{\Gamma}-\overline{X}-\overline{M}$ for comparison. (c) DFT, DFT+U (U = 3 eV), and DFT+U with SOC calculations along the Z-R-A directions. Accidental crossings protected by the nonsymmorphic symmetry in the presence of SOC along R-A are circled. (d) Calculated partial density of states from DFT. 
(e) Extracted quantities as defined in (b) for DFT+U as well as DFT+DMFT (see Supplementary Materials Supplementary Figure 8) calculations shown compared to the value from measurement, indicated by the position of the gray bar for all three axes. $E_{\alpha}$ is the bandwidth of the lower branch of the green highlighted band along Z-R; $v_F$ is its Fermi velocity; $E_{\beta}$ is the portion of the green branch along R-A below \ef.}
\label{fig:fig4}
\end{figure}

To demonstrate this protected degeneracy revealed in the above symmetry analysis, 
we first carried out DFT calculations without SOC 
(Fig.~\ref{fig:fig4}c). For illustration purposes, we show the calculated dispersions along the Z-R-A direction. Bands along additional directions are shown in the Supplementary Figure 4.
Throughout the BZ, we observe pairs of bands that disperse along Z-R and meet at the R point and then remain completely degenerate along the R-A line. One example of such pairs of bands is highlighted in green. This degeneracy exists for all bands along the BZ boundaries X-M and R-A, and is protected by the two orthogonal glide symmetries as previously discussed, resulting in a nodal line network in \pp. While DFT confirms the symmetry analysis presented earlier and show qualitative agreement with our experimental measured dispersions, we note that DFT calculated band velocities appear to be larger than that from the data, suggesting the need to investigate electronic correlations as one of the possible factors affecting the details of the electronic structure that is missing in DFT calculations. To quantify such discrepancies, we extract a number of observable quantities from the data (shown for the projected cut $\overline{\Gamma}-\overline{X}-\overline{M}$), 
including the bandwidth of the lower part of the dispersion along $\overline{\Gamma}-\overline{X}$ ($E_{\alpha}$), its Fermi velocity ($v_F$), and the portion of the hole-like dispersion along $\overline{X}-\overline{M}$ that is below \ef~($E_{\beta}$), and plot in Fig.~\ref{fig:fig4}e. It is clear that the DFT calculated bands exhibit larger bandwidth. To understand the potential role of correlation effects, we carried out DFT+U ($U = 3$~eV) calculations, which accounts for the static interactions of the Pt $d$-orbitals. While the bandwidth $E_{\alpha}$ is slightly renormalized, large deviations from measurements still remain.

Next, we consider the effect of SOC, which is expected to be substantial in \pp. When SOC is included into the DFT+U calculation (Fig.~\ref{fig:fig4}c), the degeneracy between each pair of bands along the R-A (and X-M) line is lifted. 
Nevertheless, a fourfold degeneracy is indeed retained at the high-symmetry R and A points (as well as X and M), as predicted to be protected by the nonsymmorphic symmetry. Although the above symmetry analysis only guarantees the band crossings at isolated points (X, M, R, A) on the zone boundary, it does not preclude accidental crossings of the bands between these points on the R-A  (X-M)  line. Indeed as seen in calculations with SOC, accidental crossings between pairs of bands do happen and are marked by circles in Fig.~\ref{fig:fig4}c. We emphasize that these crossings are not symmetry-enforced -- they are accidental -- but once the bands cross, the screw rotation $\hat{s}_{2y}=\{2_{010}|\frac{1}{2},0,0\}$ protects the crossing from being gapped (see the Supplementary Note 4 for more details). 

While the DFT+U calculations show qualitative agreement with our experimental observations, we note that some inconsistencies remain. \hl{One example is the bandwidth of Dirac nodal lines along the BZ boundary, which remain well above that observed experimentally (0.2~eV). Another example is the band highlighted in green along X/R to M/A in the measured dispersions shown in  Fig.~\ref{fig:fig4}b, which also has a larger bandwidth and remains above \ef~along R--A cut (Fig.~\ref{fig:fig4}c), in contrast to the data. We note that while the \kz~broadening effect could possibly explain the dispersion of this band towards \ef~(see Supplementary Note 3 and Note 6), one still needs a mechanism to reduce the bandwidth towards that experimentally observed.}
\hl{We then examine factors missing in DFT and DFT + U calculations. One possibility is the dynamic correlation effects, which we probe with DMFT by calculating} the energy-momentum-dependent spectral function (see Supplementary Note 7 for details). 
From the orbital-resolved density of states (Fig.~\ref{fig:fig4}d), the Pt $d$-orbitals and the Pb $p$-orbitals both contribute significantly near \ef. We therefore examine the effect of dynamic correlations in each.
We study the effect of the on-site Coulomb interactions ($U_{d}$) of the Pt $d$-orbitals and find that an increase from 3 eV to 8 eV causes the reduction of the bandwidth of the green-highlighted band along Z-R-A, resulting in the portion along R-A crossing \ef~(Supplementary Figure 8).
This demonstrates an overall systematic improvement with the experimental observations from DFT and DFT+U, as captured by all of the extracted quantities in Fig.~\ref{fig:fig4}e. Similar theoretical experiment on $U_{p}^{Pb}$ (while smaller than $U_{d}^{Pt}$) also shows a positive modification of this band, as demonstrated by the comparison in Fig.~\ref{fig:fig4}e.
In particular, the dispersive hole band shows an improvement over the DFT+U results in terms of its agreement with the experimental observation. \hl{However, the bandwidth of the nodal lines still remain sizeable, indicating that other effects such as orbital-dependent correlations may need to be further explored}. We note that a recent optical study also reported electron correlation effects that flatten the nodal lines in ZrSiS~\cite{BasovNatPhys}.

\subsection{Band topology}
Lastly, we note that \pp~is also classified as a strong topological insulator based on \hl{the symmetry indicator and topological quantum chemistry classification~\cite{Bradly-2017, chun-2017,Zhang2019,Tang2019, Vergniory2019,Kruthoff2017,Slager2013} in DFT calculations without $U$.} While the chemical potential crosses the bands, there exists a continuous direct gap between the valence and conduction bands buried deep (about $1$~eV) below the chemical potential in certain parts of the BZ (see Supplementary Figure 4 in the Supplementary Materials). \hl{PtPb$_4$ can thus be adiabatically transformed into an insulator without closing this gap, which protects its topological insulating nature. 
If one further considers U in a DFT + U calculation as we have done in this work, the valence and conduction bands (shown as blue and red lines in Supplementary Figure 5, respectively) will cross with a stable crossing point between $\Gamma$ and $Z$, transforming PtPb$_{4}$ into a semimetal. Despite the lack of a bulk gap in this case, the $k_{z}=0$ and $k_{z}=\pi$ planes still 
preserve time-reversal symmetry and are fully gaped. Thus, we have two 2D topological invariants ${\cal Z}_{2}$ validly defined for these two planes. ${\cal Z}_{2}=1 (0)$ at $k_{z}=0 (\pi)$ planes, indicating that the semimetal phase of PtPb$_{4}$ at $U=3$ eV is topological as well. }
\hl{We conclude that the static interaction included in the DFT + U calculation triggers a phase transition between a topological insulating state and a topological semimetal state in PtPb$_{4}$. }

In summary, we have established that \pp~is a  Dirac nodal line material which hosts band crossings protected by the nonsymmorphic symmetry of the crystal structure \hl{and narrow, nearly flat bands along the BZ boundary}. 
Moreover, we have demonstrated that the \hl{narrow bands are beyond the prediction of DFT and DFT + U calculations, which may indicate a non-negligible }orbital-dependent dynamical electron correlation \hl{or other subtle mechanism}.  PtPb$_4$ offers a platform for studying the interplay of nearly flat bands and the spin-orbit coupling against the backdrop of topological protection offered by nonsymmorphic crystalline symmetries.

\section{Methods}
\subsection{Crystal synthesis and characterization}
Single crystals of PtPb$_4$ were grown using the self-flux method with a Pt:Pb ratio of 12.5:87.5. The starting
reagents were combined in an alumina crucible and sealed in an evacuated quartz tube under a partial pressure of argon. The metals were melted and homogenized at 500 $^{\circ}$C, rapidly cooled to 360 $^{\circ}$C and then cooled at 0.5 $^{\circ}$C/hour to 300 $^{\circ}$C, at which point the crystals were separated from excess liquid flux using a centrifuge. PtPb$_4$ forms in plate-like crystals that cleave easily with size up to 1 cm. The as-grown crystals were observed to have stacking faults which could be cured by post-growth annealing, for two weeks at 250 $^{\circ}$C (see the Supplementary Note 1  and Note 2 for more details).

High resolution synchrotron powder X-ray diffraction data ($\lambda$ = 0.457861 \AA) were collected up to $2\theta = 28^{\circ}$ at the 11-BM beamline at the Advanced Photon Source (APS) of Argonne National Laboratory. Discrete detectors collected data points every 0.001$^{\circ}$ and a scan speed of 0.1$^{\circ}$ sec$^{-1}$ at room temperature. Rietveld refinement, carried out with TOPAS-Academic software, confirms that PtPb$_4$ crystallizes in the tetragonal $P4/nbm$ space group (125), as previously determined~\cite{Rosler_1951}. Attempts to refine the model in other space groups previously used for PtSn$_4$ such as $Aba2$~\cite{schubert1950kristallstruktur} or $Ccca$~\cite{kunnen2000structure} cannot account for the reflections. We also evaluated the $(1, 1, \sfrac{l}{2})$  positions and do not see additional reflections, in contrast to the report in Ref.~\cite{lee2020evidence}. 

\subsection{ARPES measurements}
ARPES measurements were carried out at beamlines 10.0.1 and 4.0.3 of the Advanced Light Source and beamline 5-2 of the Stanford Synchrotron Radiation Lightsource using a R4000, R8000 and a DA30 electron analyzer, respectively. The energy and angular resolutions were set to 20 meV and 0.3$^\circ$, respectively. An $s$-polarization geometry is used throughout. The samples were cleaved \textit{in-situ} and kept in ultra high vacuum with a base pressure lower than 4 $\times$ 10$^{-11}$ torr during measurements.

\subsection{First principle calculations}
The first-principle calculations in this work were carried out by employing the Vienna Ab initio Simulation Package (VASP) with the projector augmented wave (PAW) method~\cite{PhysRevB.59.1758}. 
We used the generalized gradient approximation (GGA), as implemented in the Perdew-Burke-Ernzerhof (PBE) functional~\cite{perdew1996}.
The cutoff parameter for the wave functions was set to be 500 eV. The Brillouin Zone (BZ) was sampled by the gamma-centered method with a k-mesh $9\times9\times9$. 
The surface states and the Wilson loop were calculated by using our in-house code $\it{TMC (\mbox{Library for \underline{T}opological \underline{M}aterial \underline{C}alculations})}$ with the iterative Green's function approach~\cite{Sancho1985} based on the maximally localized Wannier functions~\cite{Marzari1997} obtained through the VASP2WANNIER90~\cite{Mostofi2008}.
The DMFT full charge self-consistency with DFT was achieved by employing the embedded-DMFT package~\cite{Haule_Dynamical_2010}.
We first obtain the band structure using DFT with the PBE exchange correlation functional in Wien2k.
The impurity problem was solved with continuous-time quantum Monte Carlo method~\cite{Werner_Continuous_2006, Werner_Hybridization_2006, Haule_Quantum_2007, Gull_Continuous_2011}.

\section{Data Availability}

The data that support the findings of this study are available from the corresponding author upon reasonable request.


\section{Acknowledgments}
The authors acknowledge fruitful discussions with Andreas Schnyder.
This research used resources of the Advanced Light Source and the Stanford Synchrotron Radiation Lightsource, both U.S. Department Of Energy (DOE) Office of Science User Facilities under contract Nos. DE-AC02-05CH11231 and AC02-76SF00515, respectively.
Calculations were carried out at the HPC Platform of ShanghaiTech University Library and Information Services, and at School of Physical Science and Technology.
E.M. acknowledges support from U.S. Department of Energy (DOE) BES grant DE-SC0019503.
M.Y. acknowledges the support from U.S. DOE grant No. DE-SC0021421, the Robert A. Welch Foundation Grant No. C-2024, and the Gordon and Betty Moore Foundation’s EPiQS Initiative through grant no. GBMF9470. 
G.L. acknowledges the National Natural Science Foundation of China under Grant No. 11874263, the National Key R\&D Program of China under Grant No. 2017YFE0131300, and the Strategic Priority Research Program of Chinese Academy of Sciences under Grant No. XDA18010000. 
A.H.N. acknowledges the support of the U.S. NSF Grant No. DMR-1917511, and Robert A. Welch Foundation grant C-1818, as well as the hospitality of the Kavli Institute for Theoretical Physics, supported by the National Science Foundation under Grant No. PHY-1748958. A.M.H. acknowledges support from the Natural Sciences and Engineering Research Council of Canada and the CIFAR Azrieli Global Scholars program. This research was undertaken thanks in part to funding from the Canada First Research Excellence Fund, Quantum Materials and Future Technologies Program. Some of the work was supported by the U.S. DOE, Office of Science, Office of Workforce Development for Teachers and Scientists, Office of Science Graduate Student Research (SCGSR) program. The SCGSR program is administered by the Oak Ridge Institute for Science and Education for the DOE under contract number DE‐SC0014664. J.Y.C. gratefully acknowledges National Science Foundation grant No. DMR-2209804. A.W. acknowledges the support of the Eugene McDermott Graduate Fellows Program. M.Y., R.J.B. and J.S.O. acknowledge the support from NSF DMREF grants No. DMR-1921847 and No. DMR-1921798.

\section{Author Contributions}

The project was initiated by A.M.H. The single crystals were grown by A.M.H. and E.M. The ARPES measurements and analyses were carried out by H.W., J.W.H., J.S.O., R. J. B. and M.Y. with the help of S.-K.M., J.D., D.H.L., and M.H. The theoretical calculations were carried out by X.C.C., G.L. and V.L. with contributions from A.N. Powder X-ray diffraction and Rietveld refinement was carried out by A.W., G.T.M. and J.Y.C. The manuscript was written by H.W., A.M.H., A.N., G.L., E.M. and M.Y. and contributed by all the authors. A.M.H., H.W., and X.C.C. contributed equally.

\section{Competing Interests}

The authors declare no competing interests.

\bibliographystyle{naturemag}
\bibliography{bib_pp4}

\newpage

\end{document}